\documentclass[twocolumn,showpacs,prb]{revtex4}
\usepackage{graphicx}
\usepackage{dcolumn}
\usepackage{amsmath}
\usepackage{latexsym}
\begin{document}
\title{Gauge invariances vis-{\'a}-vis Diffeomorphisms in second order metric gravity:\\
 A new Hamiltonian approach}
\author{Pradip Mukherjee}
\altaffiliation{pradip@bose.res.in}
\altaffiliation{Also Visiting Associate, S. N. Bose National Centre for Basic Sciences, JD Block, Sector III, Salt Lake City, Calcutta -700 098, India and\\ IUCAA, Post Bag 4, Pune University Campus, Ganeshkhind, Pune 411 007,India}
\affiliation{Department of Physics, Presidency College,\\86/1 College Street, Kolkata-700073, West Bengal, India.}
\author{Anirban Saha}
\altaffiliation{ani\_saha09@yahoo.co.in}
\altaffiliation{Also Visiting Associate, IUCAA, Post Bag 4, Pune University Campus, Ganeshkhind, Pune 411 007,India}
\affiliation{Department of Physics, Sovarani Memorial College, Jagatballavpur, Howrah - 711 408, West Bengal, India.}

\begin{abstract}
\noindent
A new analysis of the gauge invariances and their unity with diffeomorphism invariances in second order metric gravity is presented which strictly follows Dirac's constrained Hamiltonian approach. 
\end{abstract}

\maketitle
\section{Introduction}
Einstein's General theory of relativity ( GTR ) stands as a successful theory of classical gravity which is also unique in the sense that here spacetime manifold itself acquires dynamics. 
The metric tensor $g_{\mu\nu}$ which is a measure of invariant distance between spacetime points constitute the dynamical fields of the theory.
As is well known, this feature presents great difficulties in the quantization of gravity. Many variants and extensions of GTR have been proposed which have been argued to be more suitable from one or other points of view. However, a successful theory of Quantum Gravity still eludes us \cite{QG, Carlip}. It is therefore all the more relevant to understand the classical foundations of the theories of Gravitations from different angles.

  The theories of gravitation are distinguished by a common feature which is  general covariance. From the active point of view this is the invariance of the spacetime manifold labelled by the coordinates $x^{\mu}$ under the transformations 
\begin{eqnarray}
 x^{\mu} \to x^{\prime}{}^{\mu} = x^{\mu} - \Lambda^{\mu}\left(x \right)
\label{Diff}
\end{eqnarray}
where $\Lambda^{\mu}\left(x \right)$ are arbitrary infinitesimal functions of $x^{\mu}$. This is an automorphism $M \to M$ that moves points within the manifold. Consequently there arises a certain arbitrariness of description of the gravitational field by the metric tensor 
$g_{\mu\nu}$ which can be obtained from their transformations under (\ref{Diff}). Looking from the Hamiltonian (canonical) point of view this arbitrariness is reflected in the transformations generated by the first class constraints of the theory i.e. the gauge transformations.  Stated otherwise, there should exist the right number of gauge invariances corresponding to the invariances (\ref{Diff}). The connection is however non-trivial and therefore has been a topic of continuing interest in the literature \cite{Komar, Teitelboim, Castelani,Pons1, Pons2}. 

The equivalence between the diffeomorphism (diff.) and gauge invariances is completely established when one can prescribe an exact mapping between the two sets of independent transformation parameters. While on the diff. side the independent parameters are intuitively clear, the same can not be said about the gauge parameters. Thus different works related to the subject vary not only in their interpretation of gauge transformation but also in their approach of abstracting the independent gauge parameters. As a concrete example we may consider the problem in connection with the second order metric gravity theory. In \cite{Castelani} the gauge transformations are viewed as mapping solutions to solutions and independent gauge generators are obtained following a``more Lagrangean" approach of \cite{Sudarshan} which makes use of the Lagrange equations of motion. Gauge transformations can on the otherhand be considered as mapping field configurations to field configurations.In fact this is the essence of Dirac's point of view. In \cite{Pons1, Pons2} this point of view is adopted. They find the connection between the diffeomorphism group and the gauge group by a certain projection technique from the configuration-velocity space to the phase space. Though the approaches in these works differ, they share the following common features:
\begin{enumerate}
\item All these works utilise a combination of Lagrangean and Hamiltonian methods. They can not be identified as strict Hamiltonian approaches. 
\item In one way or other these works make use of the Lagranges equations of motion. 
\end{enumerate}

In the present paper these aspects will precisely be our points of departure, i.e. our purpose here will be 
\begin{enumerate}
\item the construction of a dedicated Hamiltonian approach {\it a la} Dirac which will lead to the equivalence between the diffeomorphism and gauge transformations. 
\item to derive the most general gauge transformation generator without taking recourse to the velocity-space approach. 
\end{enumerate}
As concrete example we will also consider the second order metric gravity theory though our approach will be easily applicable to other theories of gravitation as well.

In the Canonical approach to the metric gravity a time parameter needs to be identified. This is attained by dividing space-time in to a collection of space-like three-surfaces with a time-like direction of evolution. This is the famous Arnowitt--Deser--Misner (A--D--M) decomposition \cite{ADM} where the arbitrariness of the foliation is reflected by one `lapse' and three `shift' variables. One can cast the original Einstein--Hilbert action modulo boundary terms in a form where no time derivative of these variables appear. As a consequence the corresponding momenta vanish imposing four primary constraints. Conservation of these constraints gives rise to four secondary constraints. All these constraints are first-class. Since the Hamiltonian is a linear combination of these constraints no further constraints appear. According to the Dirac conjecture the gauge generator is a linear combination of all the first-class constraints. There are thus eight gauge parameters appearing in the generator. However, only four of them are independent since the number must be equal to the number of primary first-class constraints. As has been pointed out in the above, the crucial first step in establishing a one-to-one correspondence between the diffeomorphisms and the gauge variations is to identify the independent gauge parameters. For the success of our programme (1) we need a strictly Hamiltonian method to achieve this. 

   There exists a Hamiltonian approach in the literature which provideds a general algorithm for abstracting the independent gauge parameters in any gauge theory \cite{BRR1, BRR2}. This method was applied to analyze the gauge invariances in various field and string theoretic models in the literature \cite{BRR2, ex1, ex2, us1, us2, us3}. We like to use the same algorithm here. This approach of analyzing the gauge invariances is a novel one which can be contrasted with the approach of \cite{Castelani} where a more Lagrangean approach of \cite{Sudarshan} was adopted and also with the approach of \cite{Pons1} where the gauge transformations are obtained as Legendre map from the coordinate-velocity space to the phase space. Also this algorithm is a ``dedicated" Hamiltonian algorithm in the sense that it requires only the Hamiltonian and the first-class constraints of the theory and no reference to the associated action is necessary.

 After identifying the independent gauge parameters we require to find a connection through which the gauge variations and diff. variations may be related. Again the lapse and shift variables provide this connection. There gauge variations can be immediately written down. Since they are related to the 0i-th components of the metric their variation due to reparametrization (\ref{Diff}) can be independently worked out. This will be used to establish the exact mapping between the independent gauge and diff. parameters. The mapping obtained by this connection will then be tested on the other variables to verify the consistency of the procedure. This will explicitly demonstrate the unity of the different symmetries involved. Also this mapping will enable us to compare our results with those available in the literature \cite{Castelani, Pons1, Pons2}. 

   The organization of the paper is as follows. In the next section a short review of the canonical formalism of $\left(3+1 \right)$ dimensional gravity will be given. The purpose of the review is to summarise the principal results which will be required in the sequel and also to fix the notations. Section 3 contains our analysis. We conclude in section 4.

\section{Second order canonical formalism of metric gravity}
We begin with the Einstein--Hilbert action on a manifold $M$
\begin{eqnarray}
S = \int \left(-{}^{\left(4\right)}g\right)^{1/2} {}^{\left( 4\right) }R\left(x\right) d^{4}x
\label{S} 
\end{eqnarray}
where ${}^{\left( 4\right) }R\left(x\right)$ is the Ricci scalar and ${}^{\left(4\right)}g$ is the determinant of the metric ${}^{\left(4\right)}g_{\mu\nu}$. The pre-superscript ${}^{\left(4\right)}$ indicates that the corresponding quantities are defined on the four-dimentional manifold $M$. This is required to distinguish these quantities from their analogue defined on the theree-hypersurface which are written without any such pre-superscript. 

   By adding suitable divergeces to the action (\ref{S}) we can write an equivalent Lagrangean \cite{HRT, Sunder} 
\begin{eqnarray}
\int d^{3}x {\cal{L}} =  \int d^{3}x N^{\perp} \left(g\right) ^{1/2} \left(K_{ij}K^{ij} - K^{2} + R \right) 
\label{L} 
\end{eqnarray}
where $ K = K^{i}{}_{i} = g_{ij} K^{ij}$ and $R$ is the Ricci scalar on the three surface.
The lapse variable $N^{\bot}$ represents arbitrary variation normal to the three-surface on which the state of the system are defined whereas the shift variables $N^{i}$  represent variations along the three-surface. They are defined by 
\begin{eqnarray}
N^{j} &=& g^{ij}g_{0i}
\label{Nj}
\end{eqnarray}
\begin{eqnarray}
N^{\perp} &=& \left(- g^{00}\right)^{-1/2} 
\label{N} 
\end{eqnarray}
Note that $N^{i}$ is contained in the Lagrangean (\ref{L}) through the definition of $K_{ij}$ given by
\begin{eqnarray}
K_{ij} &=& \frac{1}{2 N^{\bot}} \left(- \dot{g}_{ij} + N_{i \mid j } + N_{j \mid i} \right)
\label{K} 
\end{eqnarray}
where the ${\mid}$ indecates covariant derivative on the three-surface. 
Since the lapse and shift variables represent arbitrary deformations of the hypersurface one can expect them not to be restricted by the Hamiltonian equations. 
Hence the Lagrangean (\ref{L}) is suitable for canonical analysis because it does not contain time derivatives of $N^{\mu} \left(N^{\bot},N^{i}\right)$. One can immidiately write down the primary constraints following from the definition of the conjugate momenta of $N^{\mu}$
\begin{eqnarray}
\pi_{\mu} &=& \frac{\partial {\cal{L}}}{\partial{\dot{N}}^{\mu}} = 0
\label{M1} 
\end{eqnarray} 
The second fundamentul form of the three-surface $K_{ij}, \left(i,j = 1,2,3\right)$ contains the velocities $\dot{g}_{ij}$ and therefore related to the momenta canonical to $g_{ij}$ by 
\begin{eqnarray}
\pi^{ij} &=& \frac{\partial {\cal{L}}}{\partial{\dot{g}}_{ij}} = - \left(g\right) ^{1/2}\left(K^{ij} - K g^{ij} \right)
\label{M	2} 
\end{eqnarray} 
The inverse relation expresses $K_{ij}$ in terms of the dynamical variables of the theory
\begin{eqnarray}
K^{ij} &=&  - \left(g\right) ^{-1/2}\left(\pi^{ij} - \frac{1}{2}\pi g^{ij} \right)
\label{K1} 
\end{eqnarray} 
where $\pi = g_{ij} \pi^{ij}$.
The non-trivial Poission Brackets (PB) between the pair of conjugate variables of the theory are 
\begin{eqnarray}
\left\lbrace g_{ij} \left( x\right), \pi^{kl} \left( x^{\prime} \right)\right\rbrace &=& \frac{1}{2}\left( \delta^{i}{}_{k}\delta^{j}{}_{l} + \delta^{j}{}_{k}\delta^{i}{}_{l}\right) \delta^{\left(3 \right) }\left( x - x^{\prime}\right) \nonumber \\
\left\lbrace N^{\mu} \left( x\right), \pi_{\nu} \left( x^{\prime} \right)\right\rbrace &=& \delta^{\mu}{}_{\nu} \delta^{\left(3 \right) }\left( x - x^{\prime}\right) 
\label{PB}
\end{eqnarray} 
Using equations (\ref{L}), (\ref{K}), (\ref{M1}) and (\ref{K1}) the canonical Hamiltonian can be worked out as
\begin{eqnarray}
H_{c} &=&  \int d^{3}x \left(\pi_{\mu} \dot{N}^{\mu} + \pi^{ij} \dot{g}_{ij} - {\cal{L}}\right) \nonumber\\
&=& \int d^{3}x \left( N^{\perp}{\cal{H}}_{\perp} + N^{i}{\cal{H}}_{i} \right) 
\label{H} 
\end{eqnarray}
where,
\begin{eqnarray}
{\cal{H}}_{\perp} &=&  g ^{-1/2} \left( \pi_{ij}\pi^{ij} - \frac{1}{2} \pi^{2}\right) - \left(g\right) ^{1/2}R\\
{\cal{H}}_{i} &=& - 2 \pi_{i}{}^{j}{}_{\mid j }
\label{H2} 
\end{eqnarray}
We denote the primary constraints as 
\begin{eqnarray}
\Omega_{\mu} = \pi_{\mu}  \approx 0
\label{PC} 
\end{eqnarray}
and they are conserved with the Hamiltonian (\ref{H}) using the basic brackets (\ref{PB}) to generate the secondary constraints given by
\begin{eqnarray}
\Omega_{4} =  {\cal{H}}_{\perp} \approx 0\\
\Omega_{4+i} =  {\cal{H}}_{i} \approx 0
\label{SC} 
\end{eqnarray}
Using the basic PBs the constraint algebra becomes \cite{Dir1}
\begin{eqnarray}
\left\lbrace \Omega_{4}\left( x\right), \Omega_{4}\left( x^{\prime}\right) \right\rbrace   
&=& g^{ri} \left[ \Omega_{4+i} \left( x\right) +  \Omega_{4+i} \left( x^{\prime}\right)\right] \nonumber\\
&& \times \delta_{,i}\left(x - x^{\prime} \right) \nonumber\\
\left\lbrace\Omega_{4+i}\left( x\right), \Omega_{4}\left( x^{\prime}\right) \right\rbrace  
&=& \Omega_{4} \delta_{,i}\left(x - x^{\prime} \right) \nonumber\\
\left\lbrace\Omega_{4+i}\left( x\right), \Omega_{4+j}\left( x^{\prime}\right) \right\rbrace 
&=&  \Omega_{4+i}\left( x^{\prime}\right)\delta_{,j}\left(x - x^{\prime} \right)\nonumber\\
&& + \Omega_{4+j}\left( x\right)\delta_{,i}\left(x - x^{\prime} \right) 
\label{CA} 
\end{eqnarray}
This weakly involutive algebra  signifies that the set (\ref{PC}) - (\ref{SC}) are first-class constraints. 
This concludes our review of the canonical formulation of meric gravity. In the next section we will analyze the gauge symmetry and establish its underlying unity with the reparametrization invariance of the theory in an explicit manner.

\section{Gauge symmetry and Diffeomorphism}
We will now proceed to find the desired mapping between the independent gauge parameters and the reparametrization parameters. As mentioned in the introduction, the algorithm of \cite{BRR1, BRR2} will be followed to find the independent gauge parameters. It will thus be convenient to begin with a summary of the useful results of \cite{BRR1, BRR2}.

    Consider a theory with first class constraints only. The set of
constraints $\Omega_{a}$ is assumed to be classified as
\begin{equation}
\left[\Omega_{a}\right] = \left[\Omega_{a_1}
                ;\Omega_{a_2}\right]
\label{215}
\end{equation}
where $a_1$ belong to the set of primary and $a_2$ to the set of
secondary constraints. The total Hamiltonian is
\begin{equation}
H_{T} = H_{c} + \Sigma\lambda^{a_1}\Omega_{a_1}
\label{216}
\end{equation}
where $H_c$ is the canonical Hamiltonian and $\lambda^{a_1}$ are Lagrange multipliers enforcing the primary constraints. The most general expression for the generator of gauge transformations is obtained according to the Dirac conjecture \cite{Dir1} as
\begin{equation}
G = \Sigma \epsilon^{a}\Omega_{a}
\label{217}
\end{equation}
where $\epsilon^{a}$ are the gauge parameters. Note that all the first-class constraints appear in $G$. However, only $a_1$ of the parameters $\epsilon^{a}$ are independent, the number being equal to the number of primary first-class constraints \cite{HTZ}. By demanding the commutation of an arbitrary gauge  variation with the total time derivative,(i.e. $\frac{d}{dt}\left(\delta q \right) = \delta \left(\frac{d}{dt} q \right) $) we arrive at the following equations \cite{BRR1, BRR2}
\begin{equation}
\delta\lambda^{a_1} = \frac{d\epsilon^{a_1}}{dt}
                 -\epsilon^{a}\left(V_{a}^{a_1}
                 +\lambda^{b_1}C_{b_1a}^{a_1}\right)
                              \label{218}
\end{equation}
\begin{equation}
  0 = \frac{d\epsilon^{a_2}}{dt}
 -\epsilon^{a}\left(V_{a}^{a_2}
+\lambda^{b_1}C_{b_1a}^{a_2}\right)
\label{219}
\end{equation}
Here the coefficients $V_{a}^{a_{1}}$ and $C_{b_1a}^{a_1}$ are the structure
functions of the involutive algebra, defined as
\begin{eqnarray}
\{H_c,\Omega_{a}\} = V_{a}^b\Omega_{b}\nonumber\\
\{\Omega_{a},\Omega_{b}\} = C_{ab}^{c}\Omega_{c}
\label{2110}
\end{eqnarray}
Solving (\ref{219}) it is possible to choose $a_1$ independent
gauge parameters from the set $\epsilon^{a}$ and express $G$ of
(\ref{217}) entirely in terms of them. The other set (\ref{218})
gives the gauge variations of the Lagrange multipliers. It can be shown that
these equations are not independent conditions but appear as internal
consistency conditions. In fact the conditions (\ref{218}) follow from
(\ref{219}) \cite{BRR1, BRR2}. 

      Before proceeding further let us note the following point:\\
The assumption on which (\ref{219}) is based only involves the relation between the velocities and the canonical momenta and the arbitrary Lagrange multipliers, i.e. the first of Hamiltons equations \cite{BRR1}
\begin{eqnarray}
\dot{q} = \left[q, H_{c}\right] + \lambda^{a_{1}}\left[q, \Omega_{a_{1}}\right]
\label{q} 
\end{eqnarray}
Note that the full dynamics is not required to impose restrictions on the gauge parameters. Since this is the only input in our method of abstraction of the independent gauge parameters in the context of second order metric gravity we find that our analysis will be valid off-shell to this extent. \footnote{Of course dynamics will be needed in establishing the full equivalence of the phase space variables under the two types of transformations \cite{Carlip}.}  Also note in this connection that the equation (\ref{219}) was also obtained by an extended action procedure in \cite{HTZ}. 

For the metric gravity the gauge generator is written as 
\begin{eqnarray}
G = \int d^{3}x \left(\epsilon^{0}\Omega_{0} + \epsilon^{i}\Omega_{i} + \epsilon^{4}\Omega_{4} + \epsilon^{4+i}\Omega_{4+i}\right)
\label{GG} 
\end{eqnarray}
which is obtained from (\ref{217}) in the continum limit. The set of constraints $\Omega$ is given by (\ref{PC})-(\ref{SC}).
To find the independent gauge parameters from the set $\left(\epsilon^{0}, \epsilon^{i}, \epsilon^{4}, \epsilon^{4+i}\right)$ we require to solve the analogue of (\ref{219}), 
with the indicated parameters.
For this we require to compute the structure functions of the involutive algebra (\ref{CA}).
 
  The structure functions $C_{ab}{}^{c}$ are obtained from the second equations of (\ref{2110}) of which only $C_{b_{1}a}^{a_{2}}$ will be required in our analysis \footnote{The details of these structure functions are given in \cite{Teitelboim}}.  However, the later coefficients vanish in the present case since the primary first-class constraints $\Omega_{\mu}$ in (\ref{PC}) gives strictly zero brackets with all the constraints of the theory. The non-trivial structure factors $V_{\alpha}^{\beta}\left( x, x^{\prime}\right) $ are obtained from the first equations of (\ref{2110}) written in the continum limit as
\begin{equation}
\left\lbrace H_{c}, \Omega_{\alpha}\left( x\right) \right\rbrace   = \int d^{3}x^{\prime} V_{\alpha}^{\beta}\left( x, x^{\prime}\right) \Omega_{\beta}\left( x^{\prime}\right)
\end{equation}
Using the constraint algebra (\ref{CA}) we get 
\begin{eqnarray}
V_{4}{}^{4+s}\left(x, x^{\prime} \right) &=&  N^{\bot}\left(x^{\prime} \right) g^{rs}\left(x^{\prime} \right) \partial^{\prime}{}_{r}\delta\left(x - x^{\prime} \right) \nonumber\\
&& - \partial_{r} N^{\bot} g^{rs}\delta\left(x - x^{\prime} \right) \nonumber \\
V_{4}{}^{4}\left(x, x^{\prime} \right) &=&  N^{i}\left(x^{\prime} \right)\partial^{\prime}_{i}\delta\left(x - x^{\prime} \right)  \nonumber\\
V_{4+s}^{4}\left(x, x^{\prime} \right) &=& -\partial_{s}N^{\bot}\left(x \right)\delta\left(x - x^{\prime} \right) \nonumber \\
V_{4+s}^{4+i}\left(x, x^{\prime} \right) &=& -\partial_{s}N^{i}\delta\left(x - x^{\prime} \right) \nonumber \\
&&+ N^{l}\left(x^{\prime} \right)\partial^{\prime}_{l}\delta\left(x - x^{\prime} \right)\delta^{i}{}_{s}  \nonumber \\
V_{\mu}^{4}\left(x, x^{\prime} \right)&=&\delta^{0}{}_{\mu}\delta\left(x - x^{\prime} \right)  \nonumber \\
V_{\mu}^{4+i}\left(x, x^{\prime} \right)&=&\delta^{i}{}_{\mu}\delta\left(x - x^{\prime} \right)
\label{V} 
\end{eqnarray}

The basic equations connecting the gauge parameters (i.e. (\ref{219})) now become 
\begin{equation}
  0 = \frac{d\epsilon^{a_2}\left(x \right)} {dt}
 - \int d^{3}x^{\prime} \epsilon^{a}\left(x^{\prime} \right) V_{a}^{a_2}\left( x, x^{\prime}\right) 
\label{219a}
\end{equation} 
Using (\ref{V}) in (\ref{219a}) four equations involving the eight gauge parameters are obtained 
\begin{eqnarray}
0 &=& \left[  \dot{\epsilon}^{4} - \epsilon^{0}+ \epsilon^{4+s} \partial_{s}N^{\bot} - N^{i} \partial_{i}\epsilon^{4}\right] \left(x\right) \nonumber \\
0 &=& \left[  \dot{\epsilon}^{4+i}  - \epsilon^{i} + \epsilon^{4+s} \partial_{s}N^{i} - N^{l}\partial_{l} \epsilon^{4+i}\right. \nonumber\\
&&\left.  - N^{\bot}g^{ri}\partial_{r}\epsilon^{4} + \epsilon^{4}g^{ri}\partial_{r}N^{\bot}\right] \left(x\right)
\label{IGP1} 
\end{eqnarray}
The equations in  (\ref{IGP1}) suggest that the set $\left( \epsilon^{0}, \epsilon^{i}\right) $ will be the appropriate choice of the dependent gauge parameters. We can immediately express them in terms of remaining parameters $\left( \epsilon^{4}, \epsilon^{4+i}\right)$ as
\begin{eqnarray}
\epsilon^{0} \left(x\right) &=&\left[  \dot{\epsilon}^{4} + \epsilon^{4+s} \partial_{s}N^{\bot} - N^{i} \partial_{i}\epsilon^{4}\right] \left(x\right)\\
\epsilon^{i} \left(x\right)&=&\left[  \dot{\epsilon}^{4+i} + \epsilon^{4+s} \partial_{s}N^{i} - N^{l}\partial_{l} \epsilon^{4+i}\right. \nonumber\\
&&\left.  - N^{\bot}g^{ri}\partial_{r}\epsilon^{4} + \epsilon^{4}g^{ri}\partial_{r}N^{\bot}\right] \left(x\right)
\label{IGP} 
\end{eqnarray}
Substituting the above expressions in (\ref{GG}) we obtain the gauge generator solely in terms of the independent gauge parameters the number of which matches with the number of independent primary first-class constraints, as it should be \cite{HTZ, BRR1}. Also note that the most general form of the gauge generator contains time derivatives of the independent gauge parameters. It is remarkable that in our approach this feature follows naturally from the formalism and needs no special treatement. 

 After identifying the most general gauge generator of the theory we now proceed to derive the desired mapping between the gauge and the reparametrization parameters. This is conveniently obtained from the gauge variations of $N^{i}$, compairing them with the corresponding variations due to reparametrization (\ref{Diff}).

 The gauge variations of the shift variables are 
\begin{eqnarray}
\delta N^{i}\left(x \right) & = & \left\{N^{i}\left(x\right), G\right\}\nonumber\\  &=& \left[  \dot{\epsilon}^{4+i} + \epsilon^{4+s} \partial_{s}N^{i} - N^{l}\partial_{l} \epsilon^{4+i}\right. \nonumber\\
&&\left.  - N^{\bot}g^{ri}\partial_{r}\epsilon^{4} + \epsilon^{4}g^{ri}\partial_{r}N^{\bot}\right] \left(x\right)
\label{GV2}
\end{eqnarray}
To find the corresponding variations due to reparametrization we have to use the variations of the four-metric ${}^{\left( 4\right) }g_{\mu \nu}$ under the infinitesimal transformation (\ref{Diff})
\begin{eqnarray}
\delta  {}^{\left( 4\right) }g_{\mu \nu}  = {}^{\left( 4\right) }g_{\gamma \nu} \partial_{\mu}\Lambda^{\gamma} + {}^{\left( 4\right) }g_{\gamma \mu} \partial_{\nu}\Lambda^{\gamma} + \Lambda^{\gamma}\partial_{\gamma}  {}^{\left( 4\right) }g_{\mu \nu}
\label{gDiff}
\end{eqnarray}
Using (\ref{gDiff}) and (\ref{Nj}) we can compute the desired variations under the reparametrization (\ref{Diff}):
\begin{eqnarray}
\delta N^{i}\left(x \right) & = & \left(\frac{d}{dt} - N^{k}\partial_{k} \right) \left(  \Lambda^{i} +  \Lambda^{0}N^{i}\right) \nonumber\\
&&+ \left(  \Lambda^{k} +  \Lambda^{0}N^{k}\right) \partial_{k}N^{i} \nonumber\\
&& - \left( N^{\bot} \right)^{2} g^{ij} \partial_{i} \Lambda^{0} 
\label{R2}
\end{eqnarray}
where we have also used the inverse of the relations (\ref{N}), namely 
\begin{eqnarray}
g_{ij}N^{j} & = & N^{i} \\
g_{ij}N^{i}N^{j} - \left(N^{\bot} \right)^{2} & = & g_{00}
\label{IN}
\end{eqnarray}
Comparing the variations of the shift variable $N^{i}$ from (\ref{GV2}) and (\ref{R2}) we obtain the sought-for mapping between the reparametrization parameters and the independent gauge parameters
\begin{eqnarray}
\epsilon^{4+i} & = & \Lambda^{i} +  \Lambda^{0}N^{i}\\
\epsilon^{4} & = & N^{\bot}\Lambda^{0}
\label{M}
\end{eqnarray}
Note that similar mapping between the different sets of parameters were obtained earlier in \cite{Castelani} and also in \cite{Pons1}. Observe however that in comparison to these earlier works we follow a strictly Dirac approach of constrained Hamiltonian analysis. Moreover, we provide a structured algorithm for metric gravity where the occurence of time derivative of the gauge parameter need not be addressed seperately \cite{Castelani}. Though discussed in connection with the second order metric gravity it is apparent that this algorithm is applicable in the same general form to other theories of gravitation as well. 

    A through consistency check of the whole formalism is now in order. The mapping (\ref{M}) when used in the gauge variation of the lapse variable $N^{\bot}$
\begin{eqnarray}
\delta N^{\bot}\left(x \right) & = &\left[  \dot{\epsilon}^{4} + \epsilon^{4+s} \partial_{s}N^{\bot} - N^{i} \partial_{i}\epsilon^{4}\right] \left(x\right)
\label{GV1}
\end{eqnarray}
gives its variation in terms of the diff. parameters 
\begin{eqnarray}
\delta N^{\bot}\left(x \right) & = & \left(\frac{d}{dt} - N^{i}\partial_{i}\right)\Lambda^{0}N^{\bot} \nonumber\\
&&+ \Lambda^{0}N^{i}\partial_{i}N^{\bot} + \Lambda^{i}\partial_{i}N^{\bot}
\label{R1}
\end{eqnarray}
which is identical with the variation calculated from (\ref{gDiff}).
Similarly, we work out the gauge variation of $g_{ij}$ generated by $G$ (\ref{GG}) which gives 
\begin{eqnarray}
\delta g_{ij} \left( x \right)  & = &  \left\lbrace g_{ij} \left( x \right), G \right\rbrace \nonumber\\ 
& = & -2 \epsilon^{4} K_{ij} +  \epsilon^{4+k} \partial_{k}g_{ij} \nonumber\\
&& + g_{ki}\partial_{j} \epsilon^{4+k} + g_{kj}\partial_{i} \epsilon^{4+k} 
\label{gGV}
\end{eqnarray}
and use the mapping (\ref{M}) in it. The resulting expression can be identified with the reparametrization variation of $g_{ij}$ given by 
\begin{eqnarray}
\delta g_{ij} \left( x \right)  & = & \left( \Lambda^{0} \frac{d}{dt}  - \Lambda^{k} \partial_{k} \right) g_{ij}  + N_{i} \partial_{j} \Lambda^{0} \nonumber\\
&& + N_{j} \partial_{i} \Lambda^{0} + g_{ki}\partial_{j} \Lambda^{k} + g_{kj}\partial_{i} \Lambda^{k}
\label{gR}
\end{eqnarray}
This completes the explicit identification of the gauge invariance and diffeomorphism in second order metric gravity theory.

\section{{Conclusion}}
We discussed a novel approach of obtaining the most general gauge invariances of the second order metric gravity theory following the general Hamiltonian method of \cite{BRR1, BRR2} and used this analysis to establish a one-to-one mapping between the gauge and reparametrization parameters. We have performed explicit computation to check the consistency of our method. Though we rederive already available results \cite{Castelani, Pons1} our method is completely new in the following senses: 
\begin{enumerate}
\item This is a new dedicated Hamiltonian approach to the problem and does not require to refer to the velocity space at any stage in the calculational algorithm. As far as we know this is the first time such a calculational scheme is advanced in canonical gravity. 
\item This approach reveals properly to what extent the mapping between diffeomorphisms and gauge invariances can be considered valid off-shell. Our Hamiltonian method clearly reveals that it is dependent only on the first set of Hamilton's equations which connects the velocities, momenta and the Lagrange multipliers. In other words the specific phase space structure is only important but not the full dynamics. Note however dynamics must be invoked in establishing the equivalence of transormations of the full set of phase space variables as we have already mentioned. 
\end{enumerate}
In addition to these attractive features our method has the advantage of providing a structured algorithm which can easily be applied to other theories of gravitation.  
\section*{Acknowledgement}
The authors like to acknowledge the excellent hospitality of IUCAA where part of the work was done.


\begin{thebibliography}{99}
\bibitem{QG} Claus Kiefer {\it {Quantum Gravity}}, (Oxford University Press, 2004).
\bibitem{Komar} P.~G.~Bergmann, A.~Komar, Int. J. Theor. Phys.{\bf 5} (1972) 15.
\bibitem{Teitelboim} C.~Teitelboim, Ann.~Phys.~(N.Y.) {\bf{ 79}} (1973) 542.
\bibitem{Castelani} L.~Castellani, Ann. Phys. {\bf{143}}, (1982) 357.
\bibitem{Pons1} J.~M.~Pons, D.~C.~Salisbury and L.~C.~Shepley, Phys. Rev. {\bf D55} (1997) 658, [gr-qc/9612037].
\bibitem{Pons2} J.~M.~Pons, Class. Quant. Grav.{\bf 20} (2003) 3279, [gr-qc/0306035].
\bibitem{Sudarshan} E.~C.~G.~Sudarshan, N.~Mukunda, {\it Classical Dynamics-A Modern Perspective,}(Wiely-Interscience, 1974).
\bibitem{ADM}R.~Arnowitt, S.~Deser, C.~W.~Misner, 1962 {\it {Gravitation: an Introduction to Current Research}}, ed L.~Witten, (New York: Wiley).
\bibitem{BRR1} R.~Banerjee, H.~J.~Rothe and K.~D.~Rothe, Phys. Lett. {\bf{B 479}} (2000) 429- 438, [hep - th/9907217].
\bibitem{BRR2} R.~Banerjee, H.~J.~Rothe and K.~D.~Rothe, Phys. Lett.{\bf{B 462}} (1999) 248- 251, [hep - th/9906072].
\bibitem{ex1}S-T.~Hong, Y-W.~Kim, Y-J.~Park, K.~D.~Rothe, J. Phys. {\bf A 35} (2002) 7461, [hep-th/0204188]
\bibitem{ex2}R.~Banerjee, Phys. Rev. {\bf D 67} (2003) 105002, [hep-th/0210259]
\bibitem{us1} R.~Banerjee, P.~Mukherjee, A.~Saha, Phys. Rev. {\bf{D70}} (2004) 026006, [hep-th/0403065].
\bibitem{us2}R.~Banerjee, P.~Mukherjee, A.~Saha, Phys. Rev. {\bf{D72}} (2005) 066015, [hep-th/0501030 ]. 
\bibitem{us3}S.~Gangopadhyay, A.~Ghosh~Hazra, A.~Saha, Phys. Rev. {\bf {D74}} (2006) 125023, [hep-th/0701012 ].
\bibitem{HRT} A.~Hanson, T.~Regge, C.~Tietelboim, {\it Constrained Hamiltonian System}, (Accademia Nazionale Dei Lincei, Roma, 1976).
\bibitem {Sunder} K.~Sundermeyer, {\it {Lecture Notes in Physics 169, Constrained Dynamics}}, (Springer-Verlag, 1982).
\bibitem{Dir1} P.~A.~M.~Dirac, {\it Lectures on Quantum Mechanics}, (Yeshiva University Press, New York, 1964).
\bibitem{Carlip}S.~Carlip, Rept. Prog. Phys. {\bf 64} (2001) 885, [gr-qc/0108040] and the references therein.
\bibitem{HTZ}M.~Henneaux, C.~Teitelboim, J.~Zanelli, Nucl. Phys. {\bf B 332} (1990) 169.





\end{thebibliography}
\end{document}